# Support vector regression model for BigData systems

Alessandro Maria Rizzi

December 5, 2016

## Abstract

Nowadays Big Data are becoming more and more important. Many sectors of our economy are now guided by data-driven decision processes. Big Data and business intelligence applications are facilitated by the MapReduce programming model while, at infrastructural layer, cloud computing provides flexible and cost effective solutions for allocating on demand large clusters. In such systems, capacity allocation, which is the ability to optimally size minimal resources for achieve a certain level of performance, is a key challenge to enhance performance for MapReduce jobs and minimize cloud resource costs. In order to do so, one of the biggest challenge is to build an accurate performance model to estimate job execution time of MapReduce systems. Previous works applied simulation based models for modeling such systems. Although this approach can accurately describe the behavior of Big Data clusters, it is too computationally expensive and does not scale to large system. We try to overcome these issues by applying machine learning techniques. More precisely we focus on Support Vector Regression (SVR) which is intrinsically more robust w.r.t other techniques, like, e.g., neural networks, and less sensitive to outliers in the training set. To better investigate these benefits, we compare SVR to linear regression.

## 1 Introduction

Today BigData applications are more and more critical in our society, since they can broadly improve efficiency of enterprises and the quality of our lives. For example, according to a recent McKinsey analysis [1], BigData could have a potential impact of $300 billion to US health care only. The core of most of the BigData applications implemented today is constituted by the MapReduce programming model, which is the most common adopted solution [2]. Its open source implementation, Hadoop, can handle very large datasets [3]; according to IDC [4], in 2015 Hadoop processed half of the world data. In addition, due to a number of performance enhancements (e.g., SSD support, caching, I/O barriers elimination) implemented in Hadoop 2.x, MapReduce can fully support both interactive data analysis and traditional batch applications. Further improvements have been introduced with the Apache Tez framework, in which job

tasks are not forced to follow a strict two-phase order, but tasks can form an arbitrary complex directed-acyclic-graph and I/O barriers are totally eliminated from intermediate computations.

In this scenario, however, a challenging problem is to estimate the performance of an Hadoop (i.e., MR and Tez) job execution, i.e., predict the execution time required for completing a job. Performance prediction of BigData applications is extremely important, e.g., for correctly planning the required size a cluster (either physical or in the public cloud) must have to handle a certain workload. Traditionally, the only reliable way to predict performance has been to conduct a costly and time-consuming empirical evaluation [5]. To overcome these disadvantages, different models for performance prediction have been developed. However, modeling BigData performance is becoming more and more difficult: if we consider the recent improvements introduced by Hadoop 2.x in dynamically handling containers between map and reduce tasks or Tez DAG nodes, along with improving the overall cluster utilization, they also increased the model complexity. If we look, then, to computational paradigms newer than MapReduce, e.g., Apache Tez, modeling performance become extremely challenging. In fact there are in literature examples of simulation based analytic models (TODO REF); however, due to the number of intermediate layers involved, are extremely computationally demanding.

In this work we address the problem of performance prediction through the use of machine learning techniques. The idea is that instead of manually build a model which relates a job execution time to some of its features, machine learning techniques extract from a set of experiments this relation, which can then be used on different jobs given their features. Namely we used linear regression and Support Vector Regression (SVR) as machine learning techniques. We compared the different techniques and found that SVR performed better, since linear regression was not always applicable; nonetheless in few cases linear regression provided better results.

The rest of the paper is structured as follows: in Section 2 we briefly describe the machine learning techniques involved and the framework Hadoop and Tez. In Section 3 we explain the goals of this work and how we selected the relevant machine learning techniques features. In Section 4 we provide a description of the experiments we made for validating the approach and their results. Section 5 introduces the relevant related works. Finally, in Section 6 we draw some conclusions and present some possible future research directions.

## 2 Background

In this section we introduce the main machine learning approach used in this work and the Apache Hadoop and Tez frameworks.

## 2.1 Linear regression

Linear regression is a common statistical tool for finding a linear relation between an observed variable and a (set of) explanatory variables.

Given a set of $m$ training points $\{(\bar{x}_1, y_1), \ldots, (\bar{x}_m, y_m)\}$ where $\bar{x}_i \in \mathbb{R}^n$, $y_i \in \mathbb{R}$ are the feature vector and the target output respectively; the purpose of linear regression is to find the linear function $f(\bar{x}) : \mathbb{R}^n \to \mathbb{R}$ which minimizes the *least-square error*, i.e., $\sum_{i=1}^{m}(y_i - f(\bar{x}_i))^2$. Given that $f(\bar{x})$ is linear we have: $f(\bar{x}) = \bar{w}^T\bar{x} + b, \bar{w} \in \mathbb{R}^n$, thus obtaining the following optimization problem:

$$\min_{\bar{w},b} \quad \sum_{i=1}^{m}(y_i - \bar{w}^T\bar{x}_i - b)^2$$
$$\text{with} \quad \bar{w} \in \mathbb{R}^n, b \in \mathbb{R}$$

There is a huge amount of literature on linear regression, including several ways for efficiently computing it. Refer to specific literature, e.g., [6], for further details.

## 2.2 Support Vector Regression

Support Vector Regression (SVR) is a popular machine learning approach first described by V. Vapnik [7], famous for its robustness and insensitivity on outliers. In addition, it can scale to several dimensions.

This section is divided as follows: in Section 2.2.1 we define the basic SVR approach; in Section 2.2.2 we describe an SVR extension to handle non-linear models.

### 2.2.1 Formulation

Given a set of $m$ training points $\{(\bar{x}_1, y_1), \ldots, (\bar{x}_m, y_m)\}$ where $\bar{x}_i \in \mathbb{R}^n$, $y_i \in \mathbb{R}$ are the feature vector and the target output respectively; the purpose of SVR is to find the "flattest" linear function $f(\bar{x}) : \mathbb{R}^n \to \mathbb{R}$ which approximates every point in the training set with an error lower that $\varepsilon$, i.e., $\forall i \in [1, m].|y_i - f(\bar{x}_i)| < \varepsilon$. Since $f(\bar{x})$ is linear, $f(\bar{x}) = \bar{w}^T\bar{x} + b$, thus the *flatness* of $f(\bar{x})$ depends on vector $\bar{w} \in \mathbb{R}^n$ being small. So we can search $f(\bar{x})$ as the linear function having the smallest norm of $\bar{w}$, satisfying the constraint on the maximum error $\varepsilon$. We obtain the following optimization problem:

$$\min_{\bar{w},b} \quad \frac{1}{2}\bar{w}^T\bar{w}$$
$$\text{subject to} \quad \begin{cases} y_i - \bar{w}^T\bar{x}_i - b \leq \varepsilon \\ \bar{w}^T\bar{x}_i + b - y_i \leq \varepsilon \end{cases}$$
$$\text{with} \quad \bar{w} \in \mathbb{R}^n, b \in \mathbb{R}$$

Since function $f(\bar{x})$ could not exists if $\varepsilon$ is too small, we can relax our constraint over $\varepsilon$. We obtain the following problem:

$$\min_{\bar{w},b,\bar{\xi},\bar{\xi}^*} \quad \frac{1}{2}\bar{w}^T\bar{w} + C\sum_{i=1}^m (\xi_i + \xi_i^*)$$

$$\text{subject to} \quad \begin{cases} y_i - \bar{w}^T\bar{x}_i - b \leq \varepsilon + \xi_i \\ \bar{w}^T\bar{x}_i + b - y_i \leq \varepsilon + \xi_i^* \\ \xi_i, \xi_i^* \leq 0 \end{cases}$$

$$\text{with} \quad \bar{w} \in \mathbb{R}^n, b \in \mathbb{R}$$

In practice we add the slack variables $\xi_i, \xi_i^*$ to relax the constraint on the maximum allowed error. It is possible to tune error sensitivity through parameter $C$: the smaller is $C$ the more errors are ignored; on the contrary the bigger is $C$ the more the deviations from the points above $\varepsilon$ are considered. Although this formulation better describe SVR foundations, SVR is usually solved by considering the dual problem and reducing to a quadratic programming problem (see, e.g., [8]).

### 2.2.2 Kernels

It is also possible to make the SVR non-linear, which is applicable in case the target output has a non-linear dependency on the feature vector, on which the traditional SVR would perform poorly. The basic idea is to preprocess the feature vector through a map $\phi : \mathbb{R}^n \to \mathbb{R}^l$. The optimization problem thus becomes:

$$\min_{\bar{w},b,\bar{\xi},\bar{\xi}^*} \quad \frac{1}{2}\bar{w}^T\bar{w} + C\sum_{i=1}^m (\xi_i + \xi_i^*)$$

$$\text{subject to} \quad \begin{cases} y_i - \bar{w}^T\phi(\bar{x}_i) - b \leq \varepsilon + \xi_i \\ \bar{w}^T\phi(\bar{x}_i) + b - y_i \leq \varepsilon + \xi_i^* \\ \xi_i, \xi_i^* \leq 0 \end{cases}$$

$$\text{with} \quad \bar{w} \in \mathbb{R}^n, b \in \mathbb{R}$$

In such way it is possible to obtain a non-linear function $f(\bar{x})$. Note that the function mapping $\phi$ is usually not explicitly given, since it quickly becomes unfeasible for the large number of features.Instead it is implicitly specified through function $K(\bar{x}_i, \bar{x}_j)$ which computes the *dot* product between the given vectors, i.e., $K(\bar{x}_i, \bar{x}_j) = \phi(\bar{x}_i)^T\phi(\bar{x}_j)$. We provide in Table 1 the kernel functions which were used in our experiments. We point the reader to [8] for further details on SVR.

Table 1: Kernel functions

| Kernel | Kernel function | |
|---|---|---|
| *Linear* | $K(\bar{x}_i, \bar{x}_j) = \bar{x}_i^T \bar{x}_j$ | (1) |
| *Polynomial with degree d* | $K(\bar{x}_i, \bar{x}_j) = (\frac{\bar{x}_i^T \bar{x}_j}{n})^d$ | (2) |
| *Gaussian* | $K(\bar{x}_i, \bar{x}_j) = e^{-\frac{(\bar{x}_i - \bar{x}_j)^T (\bar{x}_i - \bar{x}_j)}{n}}$ | (3) |

## 2.3 MapReduce

MapReduce is a general computational paradigm, firstly implemented by Google in [9]. Hadoop is an open-source implementation of MapReduce. MapReduce splits the job execution into two main phases: map and reduce. The goal of MapReduce is to be able to scale to huge dataset, in a fault-tolerant way: for these reasons a MapReduce job is executed in a cluster and both phases are split into a number of tasks executed on different nodes. An additional shuffle phase is thus added after the map phase, in order to provide map intermediate results to reduce tasks and eventually relocate the computation data among different nodes.

During the *map* phase input data is partitioned, with every map task processing a partition. In fact the number of map tasks depends on input data size. The result of a map task is a set of key-value pairs which is further processed during the rest of the job. During the *shuffle* phase the output of the map is sorted to the node executing the reduce tasks according to the value of the key. Every key is fetched by only one reduce task. The *reduce* phase would take care of finalizing the computation on the key-value pairs received. Typically each reduce task would aggregate the key-value pairs received into a summary.

## 2.4 Hadoop YARN

Hadoop is the open source implementation of the MapReduce paradigm. Hadoop Yet Another Resource Negotiator (YARN) is the core of the second version of Hadoop, which completely changes Hadoop architecture, being also able to accommodate other paradigms than MapReduce.

The purpose of YARN is to manage the resource of the cluster among the different applications: instead of dedicated map and reduce slots which characterized the first version of Hadoop, resources are organized in containers which define a minimal unit of CPU and memory resources.

The assignment of containers to applications is done by the *Resource Manager* (RM), a process executed on the master node which assigns the resources according to a given scheduler, defining the optimal policy. The standard scheduler are *Capacity scheduler* and *Fair scheduler*. Capacity scheduler statically divide the cluster resources among the different organizations accessing the cluster. Fair scheduler, instead, aims to assign an equal amount of resources to different applications. In our experiments we always used Capacity scheduler.

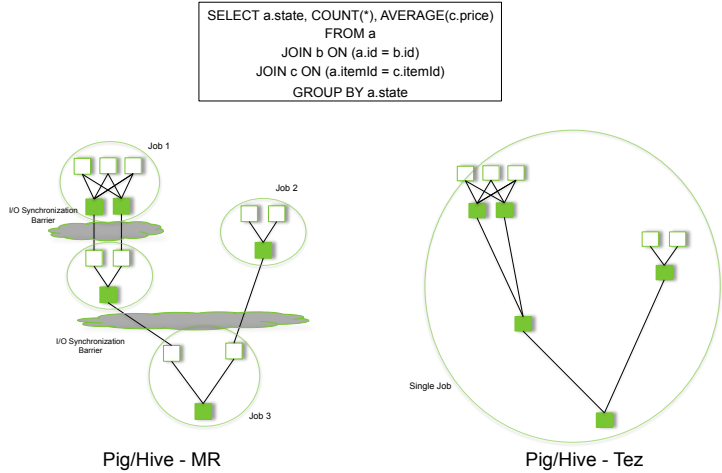

Figure 1: MR and Tez execution of a Pig/Hive query

To monitor each single application an instance of the *Application Master* (AM) is created per node. AM monitors the application state and can interact with the RM to ask more resources. Finally an instance of *Node Manager* (NM) is run on each node to monitor its state and notify the RM. NM also receives the container requests from the different AMs.

### 2.5 Apache Tez

Apache Tez is a framework which works on top of Hadoop YARN to provide a new computational paradigm. Traditional MapReduce requires the job to be structured in a phase of map task followed by a phase of reduce task. To meet this requirement, complex applications, e.g., some Hive queries, have to be split into several MapReduce jobs. Tez allows jobs constituted by an arbitrary complex directed acyclic graphs (DAGs) of map, reduce and reduce/reduce stages, which can be solved more efficiently than the translation into pure MapReduce jobs reducing I/O barriers (see, e.g., Figure 1).

## 3 Research goals

The objective of this research is to apply machine learning techniques to predict the performance of BigData job execution, in our case single MapReduce jobs and Tez jobs constituted by a DAG of tasks. The overall idea is that job execution time depends on a complex relation among some *features* describing the job and the cluster. We plan to obtain an approximation of such relation using machine learning techniques as linear regression and SVR over a set of job executions and use it to predict the execution time obtained by changing the conditions with respect to the training environment, i.e., changing the job or the cluster configuration.

First of all we want, on the basis of the relation obtained through the machine learning techniques, predict the performance of a cluster with a different number of CPU cores than the one used for training. Then we would like to use the model trained on one job to predict the execution time of a different job or the same job over a different dataset. Of course, the necessary condition to obtain good results is to choose suitable features, which can characterize the pair job-cluster. In the rest of this section we describe the features we selected along with the motivations such choice.

## 3.1 Feature selection

In this section we describe the features used to predict the job duration. We based our choice of features on the basis of previous works as [10] and [11]. In these works it is used a model for MapReduce job execution time based on the number of map and reduce tasks of the job, the average and maximum duration of map, shuffle and reduce tasks and the available number of CPU cores. We considered all these features, analyzing different choices for expressing the number of CPU cores. The reason is that in the models of [10] and [11] the execution time is inversely proportional with respect to the number of CPU cores. This is, of course, reasonable: the more computational resources are added, the more the execution time should be reduced. However, the machine learning techniques can find only linear models in the feature set, and inverse proportionality can be handled, e.g., by introducing non-linear feature. We then analyzed different candidate features for the number of CPU cores, which were considered one at a time along with the other features. These candidates have been: the plain number of CPU cores ($numCores$), the reciprocal of the number of cores ($\frac{1}{numCores}$), and a combination of the number of CPU cores with the number of map $numMap$ and reduce tasks $numReduce$ (terms included within [10] approximated formula), $\frac{numMap}{numCores}$ and $\frac{numReduce}{numCores}$.

In addition to these features justified by models presented in literature, we also considered other ones. In particular we introduced some measures of the amount of data involved in the job: a descriptor of the data size involved in the job and the average and maximum number of bytes transferred during shuffles.

All the features we have discussed so far have been used for pure two stage MapReduce jobs. For what concerns Tez jobs, we considered the same features we discussed up to now, considering the map, shuffle and reduce average and maximum time of the task for each stage of the job DAG. This choice in fact simply generalize the approach taken for MapReduce jobs, however, it does not allow to generalize the trained model to jobs with different DAGs.

```
select avg(ws_quantity),
avg(ws_ext_sales_price),
avg(ws_ext_wholesale_cost),
sum(ws_ext_wholesale_cost)
from web_sales
where (web_sales.
    ws_sales_price
between 100.00 and 150.00)
or (web_sales.ws_net_profit
between 100 and 200)
group by ws_web_page_sk
limit 100;
```

(a) R1

```
select inv_item_sk,
    inv_warehouse_sk
from inventory
where inv_quantity_on_hand >
    10
group by inv_item_sk,
    inv_warehouse_sk
having sum(
    inv_quantity_on_hand) > 20
limit 100;
```

(b) R2

```
select avg(ss_quantity),
avg(ss_net_profit)
from store_sales
where ss_quantity > 10
and ss_net_profit > 0
group by ss_store_sk
having avg(ss_quantity) > 20
limit 100;
```

(c) R3

```
select cs_item_sk, avg(
    cs_quantity) as aq
from catalog_sales
where cs_quantity > 2
group by cs_item_sk;
```

(d) R4

```
select inv_warehouse_sk,
sum(inv_quantity_on_hand)
from inventory
group by inv_warehouse_sk
having sum(
    inv_quantity_on_hand) > 5
limit 100;
```

(e) R5

Figure 2: MapReduce Hive queries

## 3.2 Machine learning techniques selection

We apply the learning of BigData jobs execution time dependency from the previously mentioned features using different approaches, involving linear regression and SVR. We used linear regressing as a baseline; we applied SVR both with and without kernels (Equation 1, see Table 1). The kernels we considered are the polynomial (Equation 2) with degree 2,3,4,6 and Gaussian (Equation 3).

## 4 Experimental evaluation

We now analyze and compare linear regression and SVR. The section is organized as follows: in Section 4.1 we describe the benchmark used and the clusters on which the benchmark has been executed. In Section 4.2, instead, we present and analyze the obtained results.

```sql
select a.aq from (select cs_item_sk, avg(cs_quantity) as aq
from catalog_sales
where cs_quantity > 2
group by cs_item_sk) a join (select i_item_sk, i_current_price
        from item
        where i_current_price > 2
        order by i_current_price)
        b on a.cs_item_sk = b.i_item_sk
order by a.aq
limit 100;
```

(a) Q2

```sql
select avg(ws_quantity)
,avg(ws_ext_sales_price)
,avg(ws_ext_wholesale_cost)
,sum(ws_ext_wholesale_cost)
from web_sales
where (web_sales.
    ws_sales_price between
    100.00 and 150.00) or (
    web_sales.ws_net_profit
    between 100 and 200)
limit 100;
```

(b) Q3

```sql
select inv_item_sk,
    inv_warehouse_sk
from inventory where
    inv_quantity_on_hand > 10
group by inv_item_sk,
    inv_warehouse_sk
having sum(
    inv_quantity_on_hand)>20
order by inv_warehouse_sk
limit 100;
```

(c) Q4

Figure 3: Tez Hive queries

## 4.1 Experimental setup

We considered as a benchmark different Hive queries applied on the TPC-DS[1] dataset. TPC-DS ha been chosen since it is the reference benchmark for data warehousing. We used queries R1-5, shown in Figure 2, as MapReduce jobs; and queries Q2-4, shown in Figure 3, as Apache Tez jobs. We applied them onto different datasets ranging from 250GB to 1TB for R1-5 and ranging from 40GB to 50GB for Q2-4.

All our MapReduce experiments were run on CINECA, the Italian supercomputing center, on the Big Data cluster PICO[2]. The experiments involving Apache Tez were instead performed on a dedicated cluster on Flexiant[3].

### 4.1.1 PICO cluster

PICO is composed of 74 nodes, each of them having two Intel Xeon 10 Core 2670 v2@2.5GHz, and 128GB of memory per node. Of this 74 nodes, up to 66 are available for computation. In our experiments on PICO, we used several configurations ranging from 40 to 120 cores and set up the scheduler to provide one container per core.

---

[1] http://www.tpc.org/tpcds/
[2] http://www.hpc.cineca.it/hardware/pico
[3] https://www.flexiant.com/

The cluster is shared among different users, it provides the Portable Batch System (PBS) PBS Professional to submit jobs and check their execution: it is possible to schedule a given job for execution on a given number of nodes using each a certain number of CPU cores and memory. Since the cluster is shared among different users, the performance of the execution of the single job depends on the general load of the system, even though the PBS try to split the resources among the different users. This especially affects the storage performance, since it is not handled directly by the PBS and for a large part of it is separated from the local computation nodes. Due to this, it is possible to have large variations in the performance of the class, according to the total usage of the cluster.

We tried to mitigate this variability first of all by requesting entire nodes of the cluster for the execution of our experiments. In such way, we could reasonably be sure that nobody else could run other jobs on those nodes and thus interfering to the performance measurement. Every experiment is run into an ephemeral Hadoop cluster built into the selected node, which is created at the beginning of the experiment. The PICO cluster provides the myHadoop tool for setting up an Hadoop 2.5.1 cluster, upon which we used Hive version 1.2.1. The HDFS storage is kept locally on the selected node, which diminishes the amount of variability with respect to the centralized storage. In spite of these precautions, still the experiment shown high variability, in particular with few runs characterized by extremely high execution time. To further reduce this, we discarded from our analysis experimental runs which behave significantly different. The criteria we used consist in discarding all the experiments with a difference from the average larger than 3 times the standard deviation.

### 4.1.2 Cluster Flexiant

The cluster Flexiant is part of an IaaS public cloud. It is composed of 7 nodes, of which 2 are master nodes and 5 slave (computational) nodes. Each node has 4 CPU cores and 8GB of memory. The cluster is provided with Hortonworks Data Platform (HDP) 2.3, which includes Hadoop 2.7.1 and Hive 1.2.1. Again we kept the HDFS storage local to each node to reduce variability.

## 4.2 Experimental result

In this section we describe the experiments we conducted to assess the accuracy of machine learning approaches to predict job duration.

The purpose of this section is to assess the performance of machine learning techniques to predict BigData job execution time. First we check that the feature we defined are adequate by predicting the execution of jobs equals to the ones used for training. Then we check how the machine learning techniques can predict the execution time for a different number of CPU cores or a different query.

Each experiment has been conducted in the following way: the relevant data has been partitioned in three parts: a *training set*, a *cross-validation* set and a testing set. The machine learning approach is trained on the training set.

Table 2: Mean error for query validation

| Model | R1 | R2 | R3 | R4 | R5 |
|---|---|---|---|---|---|
| Linear regression | 5.59% | —% | 1.64% | 2.23% | 2.48% |
| Linear SVR | 8.37% | 36.25% | 5.11% | 3.80% | 2.38% |
| Polynomial SVR (2) | 64.55% | 32.98% | 35.27% | 19.11% | 3.18% |
| Polynomial SVR (3) | 20.49% | 70.51% | 17.26% | 9.91% | 3.28% |
| Polynomial SVR (4) | 55.94% | 68.62% | 53.27% | 15.63% | 4.21% |
| Polynomial SVR (6) | 175.85% | 71.64% | 74.38% | 30.27% | 5.20% |
| Gaussian SVR | 11.43% | 47.65% | 9.21% | 6.35% | 2.99% |

The cross-validation set is used to find the parameters of the machine learning approach, e.g., $C$ and $\epsilon$ of Section 2.2.1. In fact, the same machine learning approach is trained multiple times with different parameters on the training set, among which the one giving the minimal error on the cross-validation set is selection. The testing set, instead, is used to compute the accuracy on the previously selected model. Note that from now on we identify as training data both training and cross-validation set.

The rest of the section is divided as follows: in Sections 4.2.1–4.2.4 we analyze MapReduce jobs; in Sections 4.2.5–4.2.6 Tez ones. First we check the effectiveness of our approach using the same job for training and testing (Section 4.2.1 for MapReduce and Section 4.2.5 for Tez). Then we try to predict the execution time of a job using a different number of CPU cores (Section 4.2.2 for MapReduce and Section 4.2.6 for Tez).

For MapReduce we also try to predict the execution of a different job in Section 4.2.3 (since in case of Tez is only possible if the jobs involved share the same DAG). In Section 4.2.4 we analyze the weight of the features involved for the MapReduce case.

### 4.2.1 MapReduce query validation

In this section we describe a preliminary experiment to validate our approach: we use the data of each query run for both training and testing, randomly split as follows: 80% training data (60% for training and 20% for cross-validation) and 20% testing data.

Since in this first experiment the training query is the same for testing, we expect to obtain good predictions. We analyze separately the case in which the number of core has been captured by feature $numCores$, $\frac{1}{numCores}$.

In the first case, shown in Table 2, we obtained fairly good result, except for query R2. Regarding the different machine learning techniques, linear regression and linear SVR behave very well, followed by Gaussian SVR and polynomial SVR. The worst average error on a single query obtained is 5.59%, 8.37%, 11.43% for linear regression, linear SVR and Gaussian SVR respectively, in query R1. Query R2 has the worst results of the five: linear SVR has an average error of 36.25%, Gaussian SVR of 47.65%. This is not surprising since R2 is the fastest query of the set and also the most affected by the noise.

Table 3: Mean error for query validation using $\frac{1}{nCores}$

| Model | R1 | R2 | R3 | R4 | R5 |
|---|---|---|---|---|---|
| Linear regression | 5.48% | —% | 2.01% | 2.16% | 3.18% |
| Linear SVR | 10.22% | 53.10% | 8.87% | 3.82% | 1.30% |
| Polynomial SVR (2) | 28.28% | 45.41% | 32.35% | 14.12% | 2.69% |
| Polynomial SVR (3) | 19.67% | 72.85% | 17.93% | 10.24% | 2.85% |
| Polynomial SVR (4) | 51.72% | 69.33% | 55.77% | 12.15% | 3.96% |
| Polynomial SVR (6) | 55.24% | 71.53% | 27.97% | 19.64% | 5.61% |
| Gaussian SVR | 11.32% | 45.34% | 8.69% | 6.85% | 2.05% |

In the second case, shown in Table 3, we obtained similar results: except R2 the worst average error of the queries is on R1 of 5.48%, 10.22%, 11.32% for linear regression, linear SVR and Gaussian SVR respectively. We obtain even worse results for query R2: the best relative error is 45.34% obtained with the Gaussian SVR.

As expected, we obtained on most of the experiments quite good results, which confirm the feasibility of our approach. Although the linear regression seems to provide the best results, in some cases, denoted with "—" in Tables 2 and 3, is not applicable.

### 4.2.2 MapReduce prediction varying CPU cores

As the first objective of this work, we want to predict the execution time on a cluster with a different number of cores than the one used for training. Basically we remove from the training set the data related to a particular number of CPU cores and use the data on the others for training. This task can be classified as *extrapolation* or *interpolation* according to the location of the predicted number of cores with respect to the training data: we have extrapolation when the predicted value is outside the range of training data; interpolation otherwise. We now focus only on feature $\frac{1}{nCores}$ for expressing the number of CPU cores, which provides better results since there is an inverse proportionality between the execution time and the number of CPU cores.

Tables 4–7 contain a summary of the obtained results; each table contains the mean relative error (in percentage) for the prediction on a number of 60, 80, 100, 120 respectively. Predictions for 120 cores are extrapolations; for 60, 80, 100 cores are interpolations.

The columns contain the different queries; the rows different machine learning techniques. For almost every conditions except with query R2 the best techniques has been linear regression, when applicable, followed by linear SVR, Gaussian SVR and polynomial SVR. However linear regression, as previously seen, is unstable and not always applicable (see the cell containing "—").

Table 4: Mean error for 60 cores predictions

| Model | R1 | R2 | R3 | R4 | R5 |
|---|---|---|---|---|---|
| Linear regression | 6.52% | 3.04% | 6.23% | 5.51% | 5.20% |
| Linear SVR | 7.53% | 10.69% | 6.68% | 14.81% | 3.90% |
| Polynomial SVR (2) | 30.04% | 23.62% | 29.45% | 45.27% | 11.99% |
| Polynomial SVR (3) | 20.60% | 71.09% | 21.96% | 35.34% | 32.46% |
| Polynomial SVR (4) | 34.42% | 69.01% | 26.35% | 63.45% | 28.81% |
| Gaussian SVR | 16.84% | 19.67% | 15.45% | 26.21% | 10.17% |

Table 5: Mean error for 80 cores predictions

| Model | R1 | R2 | R3 | R4 | R5 |
|---|---|---|---|---|---|
| Linear regression | —% | —% | 2.21% | —% | 18.38% |
| Linear SVR | 11.21% | 13.86% | 6.10% | 2.99% | 2.34% |
| Polynomial SVR (2) | 37.27% | 48.68% | 52.65% | 31.47% | 4.49% |
| Polynomial SVR (3) | 45.46% | 78.57% | 32.89% | 20.04% | 4.16% |
| Polynomial SVR (4) | 28.84% | 79.18% | 41.52% | 53.71% | 5.39% |
| Gaussian SVR | 10.08% | 34.82% | 11.43% | 6.62% | 3.28% |

If we do not consider R2, For extrapolation of 120 cores the best technique is linear SVR, with a worst average error observed in R2, of 24.85%. For interpolation of 60–100 cores, with the excpet of query R2 on 100 cores, linear SVR maintain the average error under 15%. Linear regression, when applicable, also performs quite well; with the largest error of 23.36%.

### 4.2.3 MapReduce prediction varying Hive queries

We then applied machine learning approaches to predict the execution time of one query based on the data of a different one. The training (and cross-validation) data are composed by all the available runs for any number of cores and dataset size of one query. The testing set, instead, is composed of the run for any number of CPU cores and dataset size of the query to be predicted.

Table 6: Mean error for 100 cores predictions

| Model | R1 | R2 | R3 | R4 | R5 |
|---|---|---|---|---|---|
| Linear regression | 23.36% | 17.96% | 3.49% | —% | 5.39% |
| Linear SVR | 12.62% | 72.93% | 10.71% | 6.12% | 2.67% |
| Polynomial SVR (2) | 91.52% | 26.42% | 41.28% | 35.50% | 4.17% |
| Polynomial SVR (3) | 40.88% | 70.48% | 26.57% | 19.87% | 3.18% |
| Polynomial SVR (4) | 196.02% | 70.35% | 39.79% | 29.78% | 3.80% |
| Gaussian SVR | 23.77% | 56.55% | 18.72% | 11.16% | 2.79% |

Table 7: Mean error for 120 cores predictions

| Model | R1 | R2 | R3 | R4 | R5 |
|---|---|---|---|---|---|
| Linear regression | 49.67% | 2.89% | 10.00% | 11.83% | 34.09% |
| Linear SVR | 8.93% | 24.85% | 22.33% | 8.03% | 2.83% |
| Polynomial SVR (2) | 61.33% | 53.23% | 65.48% | 78.85% | 5.37% |
| Polynomial SVR (3) | 123.09% | 73.79% | 55.10% | 48.98% | 4.35% |
| Polynomial SVR (4) | 113.89% | 70.42% | 59.48% | 103.42% | 4.47% |
| Gaussian SVR | 50.46% | 107.83% | 56.34% | 30.66% | 3.03% |

Table 8: Mean error for query-query prediction with $nCores$

| Model | R5→R2 | R2→R5 | R3→R4 | R4→R3 |
|---|---|---|---|---|
| Linear regression | 7.67% | 42.11% | 8.87% | 10.39% |
| Linear SVR | 46.76% | 278.68% | 18.31% | 10.71% |
| Polynomial SVR (2) | 46.76% | 390.74% | 139.63% | 60.33% |
| Polynomial SVR (3) | 46.76% | 319.74% | 126.14% | 46.56% |
| Gaussian SVR | 46.76% | 2781.36% | 52.76% | 27.27% |

In this case we focus on *numCores* results, which are shown in Tables 8 and 9. Tables 10 and 11, intead, show the results for $\frac{1}{numCores}$.

First we use for training and test similar queries. We start by predicting R2 from R5 and vice-versa. In both case, we obtain mixed result; the technique providing best results is linear regression, followed by linear SVR, Gaussian SVR and polynomial SVR. For linear regression we have an error of 7.67% for predicting R2 from R5, and 42.11% when predicting R5 from R2. This is not very surprising since these queries, especially R2, are characterized by small execution time and high noise, as we have previously seen. If we consider R3 and R4, we obtain much better results: an error of 8.87%, 18.31% for linear regression and linear SVR respectively when predicting R4 from R3; an error of 10.39%,10.71% respectively the other way around. Again the order of the machine leaning techniques performance is the same, with linear regression and linear SVR dominating and polynomial SVR at the end.

In both previous cases we trained and tested either two fast queries or two slow ones. Now, we validate the case in which we train on a combination of fast-slow queries and test an intermediate one. In practice we predict R3 on the training over R1,R2, and R4. The results are pretty good: the average error is 5.64% and 6.12% for linear regression and linear SVR respectively; again the rank of performances among the machine learning techniques is the same.

The last evaluation we attempted has been to mix different queries, which is predict R3 from R2 and R5 and predict R2 from R3 and R4. In this case the result were pretty bad, in a way remarking the fact that the queries considered are too different to be generalized.

Table 9: Mean error for multiquery-query prediction with $nCores$

| Model | R1,R2,R4→R3 | R2,R5→R3 | R3,R4→R2 |
|---|---|---|---|
| Linear regression | 5.64% | 86.89% | 154.94% |
| Linear SVR | 6.12% | 73.15% | 68.54% |
| Polynomial SVR (2) | 46.12% | 81.33% | 1784.38% |
| Polynomial SVR (3) | 36.92% | 80.25% | 1111.44% |
| Gaussian SVR | 11.92% | 59.76% | 811.19% |

Table 10: Mean error for query-query prediction with $\frac{1}{nCores}$

| Model | R5→R2 | R2→R5 | R3→R4 | R4→R3 |
|---|---|---|---|---|
| Linear regression | 8.30% | 43.30% | 8.66% | 11.33% |
| Linear SVR | 46.76% | 343.38% | 16.45% | 12.59% |
| Polynomial SVR (2) | 46.76% | 430.90% | 95.42% | 65.41% |
| Polynomial SVR (3) | 46.76% | 323.27% | 170.83% | 36.24% |
| Gaussian SVR | 46.76% | 2741.39% | 51.27% | 24.87% |

#### 4.2.4 MapReduce weight of the selected features

In this section we analyze the importance each of the selected features has in the prediction of the query execution time. We can see that analyzing the weight each feature has in the execution time estimation. Table 12 shows the weights considering for the cluster size either $numCore$ or $\frac{1}{numCore}$ for all the R1-5 queries. If we analyze all the features, considering the number of core as feature $numCore$, the dominant feature is the max shuffle time, which has a weight of 0.9355. Then, there are the average number of bytes transferred, with a weight of 0.1972; the average time of a reduce task, weight 0.1413. All the other features have weights less than 0.06.

If we switch feature $numCore$ with its inverse $\frac{1}{numCore}$, we obtain similar results: again the feature with highest weight is the maximum shuffle task time (0.9327), followed by the average number of bytes transferred and the average time of a reduce task, with weights -0.1550 and 0.1376 respectively.

Table 11: Mean error for multiquery-query prediction with $\frac{1}{nCores}$

| Model | R1,R2,R4→R3 | R2,R5→R3 | R3,R4→R2 |
|---|---|---|---|
| Linear regression | 5.84% | 87.05% | 190.06% |
| Linear SVR | 6.10% | 73.44% | 97.69% |
| Polynomial SVR (2) | 44.96% | 81.40% | 1670.06% |
| Polynomial SVR (3) | 33.68% | 80.25% | 1275.93% |
| Gaussian SVR | 12.61% | 62.68% | 826.95% |

Table 12: Feature weight

| Feature | $nCores$ | $\frac{1}{nCores}$ |
|---:|---:|---:|
| Map tasks | 0.029655 | 0.059395 |
| Reduce tasks | −0.025907 | −0.053398 |
| Average map task duration | 0.022160 | 0.029537 |
| Maximum map task duration | 0.013312 | 0.009639 |
| Average reduce task duration | 0.141263 | 0.137604 |
| Maximum reduce task duration | 0.002008 | 0.008253 |
| Average shuffle task duration | −0.025151 | −0.025066 |
| Maximum shuffle task duration | 0.935545 | 0.932685 |
| Average bytes transferred per shuffle task | −0.197170 | −0.154967 |
| Maximum bytes transferred per shuffle task | 0.056686 | 0.007358 |
| Dataset size | 0.015942 | 0.012835 |
| Number of CPU cores | −0.024642 | 0.020706 |

Table 13: Mean error for query validation

| Model | Q2 | Q3 | Q4 |
|---:|---:|---:|---:|
| Linear regression | 1.71% | 2.67% | 4.90% |
| Linear SVR | 1.63% | 5.77% | 4.15% |
| Polynomial SVR (2) | 11.52% | 19.05% | 16.63% |
| Polynomial SVR (3) | 5.14% | 7.83% | 7.02% |
| Polynomial SVR (4) | 11.39% | 14.70% | 12.80% |
| Polynomial SVR (6) | 11.58% | 14.65% | 14.98% |
| Gaussian SVR | 2.56% | 3.00% | 3.80% |

By analyzing the weights for each query individually, we note that the maximum shuffle task time has an high weight especially for long queries, e.g., R1 (0.8306) and R3 (0.9866). The fast queries (R2, R5) shows instead different behaviors: R2 is strongly dependent on the maximum time of a map task (0.8010); R5 on average and maximum bytes transferred per shuffle task(-0.5889 and 0.6072), followed by average map task time and number of map and reduce tasks (0.4805, 0.3576, 0.4186).

#### 4.2.5 Tez query validation

In this section we show the result of the same validation described in Section 4.2.1 to Tez queries Q2-4. Again we use the data of each query run for both training and testing: 80% training data (60% for training and 20% for cross-validation) and 20% testing data. We focused on the 50GB dataset for every query Q2, Q3 and Q4.

The results are shown in Tables 13 and 14. In this case the results are pretty good, the mean error is under 7% in every case with the exception of polynomial

Table 14: Mean error for query validation using $\frac{1}{nCores}$

| Model | Q2 | Q3 | Q4 |
|---|---|---|---|
| Linear regression | 2.88% | 2.67% | 5.98% |
| Linear SVR | 2.86% | 6.85% | 6.27% |
| Polynomial SVR (2) | 9.66% | 19.16% | 13.27% |
| Polynomial SVR (3) | 6.71% | 9.03% | 7.34% |
| Polynomial SVR (4) | 9.98% | 15.42% | 10.81% |
| Polynomial SVR (6) | 11.43% | 15.49% | 11.58% |
| Gaussian SVR | 2.94% | 2.85% | 4.03% |

Table 15: Mean error for 8 cores predictions

| Model | Q2 | Q3 | Q4 |
|---|---|---|---|
| Linear regression | 19.96% | —% | 38.24% |
| Linear SVR | 14.58% | 26.21% | 31.33% |
| Polynomial SVR (2) | 37.95% | 69.80% | 60.60% |
| Polynomial SVR (3) | 13.41% | 70.69% | 10.92% |
| Gaussian SVR | 13.71% | 9.09% | 14.71% |

SVR for both $nCores$ and $\frac{1}{nCores}$. The best machine learning techniques are linear SVR, Gaussian SVR and linear regression which have comparable performance, followed by polynomial SVR.

### 4.2.6 Tez prediction varying CPU cores

We now repeat for Tez queries the analysis shown in Section 4.2.2 for MapReduce queries. In this case we try to use the data gathered from all the cluster size except one for training and then use the learned model to predict the execution time of the missing size. In this case, we applied this procedure for cores 8, 12, 16 and 20. As before, 8 and 20 cores predictions are cases of extrapolation, since the size is outside the range of the available dataset; 12 and 16, instead, are cases of interpolation. We thus expect better results for 12 and 16 cores, with respect to 8 and 20. We show here, in Tables 15–18, the obtained results using $\frac{1}{nCores}$. We obtained the best results for 12 cores, in which the mean error of linear regression, linear SVR and Gaussian SVR is always lower than 17%. For 16 cores, only Gaussian SVR produce acceptable results, with mean error less than 21% for every condition. The case with 20 cores gives the worst result; Gaussian SVR produce has mean error less than 43% for every condition. Finally for 8 cores, again the Gaussian SVR yields the best result, with a mean error under 15% in every case.

Table 16: Mean error for 12 cores predictions

| Model | Q2 | Q3 | Q4 |
|---|---|---|---|
| Linear regression | 9.53% | 16.35% | 15.57% |
| Linear SVR | 7.31% | 15.82% | 14.70% |
| Polynomial SVR (2) | 16.69% | 35.73% | 30.76% |
| Polynomial SVR (3) | 8.28% | 20.39% | 19.58% |
| Gaussian SVR | 6.06% | 13.17% | 14.95% |

Table 17: Mean error for 16 cores predictions

| Model | Q2 | Q3 | Q4 |
|---|---|---|---|
| Linear regression | 9.75% | —% | 48.12% |
| Linear SVR | 8.38% | 49.56% | 37.84% |
| Polynomial SVR (2) | 14.04% | 46.57% | 40.01% |
| Polynomial SVR (3) | 11.73% | 16.60% | 23.54% |
| Gaussian SVR | 9.47% | 8.16% | 20.52% |

# 5 Related work

In the field of computer systems performance modeling many works adopt machine learning techniques. In [12], multilayer neural networks are used to model high performance parallel applications. In [13], instead, artificial neural networks are used for obtaining models of virtualized application environments; in [14] CPU and memory resource provisioning is evaluated using regression models; in particular simple neural networks and M5' trees are considered. In the scenario of data centers performance tuning, [15] uses different machine learning techniques (including neural network and SVM) to configure memory prefetchers. [16] adopts artificial neural networks to predict impact on performance of architectural changes; similarly [17] predicts performance and power of a microprocessor configuration using regression. Related to the field of "Green" IT, [18] adopts linear regression and M5P decision tree algorithm to predict power consumption, CPU loads, and SLA timings.

Table 18: Mean error for 20 cores predictions

| Model | Q2 | Q3 | Q4 |
|---|---|---|---|
| Linear regression | 3.46% | 43.38% | 116.00% |
| Linear SVR | 3.10% | 41.85% | 107.14% |
| Polynomial SVR (2) | 19.08% | 61.06% | 103.09% |
| Polynomial SVR (3) | 10.18% | 39.69% | 23.52% |
| Gaussian SVR | 4.01% | 22.14% | 42.43% |

For what concerns MapReduce applications performance modeling, one of the most explored directions consists in providing approximated formulas devised to capture the framework internal mechanisms and job properties. For instance, [19] obtain in this way a parametric model to predict performance given the cluster configuration and application-specific data.

Building upon similar analytical models, [20] take into account the detrimental effects of resource contention and task failures to achieve a more precise prediction of completion times. [21], instead, explore the impact on performance of exploiting heterogeneous computational nodes in Cloud-based Hadoop clusters, allowing for tailoring resources to the needs of specific applications.

More recently, as in our work, machine learning techniques are used to support performance prediction of Hadoop clusters. [22] compare through an experimental campaign several methods, ranging from ordinary linear regression to advanced techniques as neural networks, model trees, and support vector regression (SVR). In most experiments SVR shows the best accuracy, with only one case where neural networks perform better.

SVR has also been adopted also to implement automated resource allocation and configuration for Cloud-based clusters, as proposed by [23]. The described AROMA system mines historical execution data in order to profile past submissions, then it robustly matches incoming jobs to the available performance signatures for prediction. In this way, AROMA allows for meeting deadlines stated in Service Level Agreements (SLAs) incurring minimum cost, with an average accuracy on completion time estimates around 12%.

Authors in [24] proposed a methodology and a framework based on multivariate linear regression to predict the execution time of iterative algorithms for the Apache Giraph implementation able to achieve a relative prediction error around 10%-30%. However, the approach is very specific to predict the performance of DAGs analyzing homogeneous graph structures with a global convergence condition (e.g., an accuracy threshold for the estimation of a graph metric).

With respect to previous works, we applied thoroughly different kind of SVR, comparing the obtained results. More precisely, we applied both black-box approaches obtained through Gaussian and polynomial kernel application, to a gray box model, obtained mixing MapReduce models, which defines the interesting features, to machine learning techniques as linear regression and SVR.

# 6  Conclusion and future work

The application of machine learning techniques to BigData job execution time prediction can provide precise predictions in many situations. In particular, we obtained promising results in predicting the execution time by varying the cluster size. With respect to performance prediction of a different job, the results depended of the similarity between the jobs. It is also important to remark that some results were biased by some noise affecting especially some queries, e.g., R5 and R2. Overall SVR provided better results than linear regression, since in different situation the latter does not converge and only in few samples provides better accuracy. The best results have been obtained considering the feature $\frac{1}{nCores}$, which performs better than any non-linear support vector regression (polynomial and Gaussian).

Possible future work include the application of other machine learning algorithm (e.g., neural networks) or the analysis on different clusters or with other kind of BigData job (e.g., Spark[4] jobs). Promising directions involve the combination of analytical models (e.g., queuing networks) and machine learning techniques, e.g. reference [25]: through the use of existing expensive models it is possible to generate lots of data which can be fed to a machine learning techniques to extract simpler relations, which can be further improved by adding real data from BigData job execution in a cluster.

# References

[1] J. Manyika, M. Chui, B. Brown, J. Bughin, R. Dobbs, C. Roxburgh, and A. H. Byers, *Big Data: The next frontier for innovation, competition, and productivity*, McKinsey Global Institute, 2012.

[2] K.-H. Lee, Y.-J. Lee, H. Choi, Y. D. Chung, and B. Moon, "Parallel data processing with MapReduce: A survey," *SIGMOD Rec.*, vol. 40, no. 4, pp. 11–20, Jan. 2012, ISSN: 0163-5808. DOI: `10.1145/2094114.2094118`.

[3] F. Yan, L. Cherkasova, Z. Zhang, and E. Smirni, "Optimizing power and performance trade-offs of MapReduce job processing with heterogeneous multi-core processors," in *CLOUD*, 2014. DOI: `10.1109/CLOUD.2014.41`.

[4] K. Kambatla, G. Kollias, V. Kumar, and A. Grama, "Trends in Big Data analytics," *Journal of Parallel and Distributed Computing*, vol. 74, no. 7, pp. 2561–2573, 2014, ISSN: 0743-7315. DOI: `10.1016/j.jpdc.2014.01.003`.

[5] G. P. Gibilisco, M. Li, L. Zhang, and D. Ardagna, "Stage aware performance modeling of DAG based in memory analytic platforms," in *Cloud*, 2016.

[6] G. A. Seber and A. J. Lee, *Linear regression analysis*. John Wiley & Sons, 2012, vol. 936.

[7] V. N. Vapnik, *The nature of statistical learning theory*. New York, NY, USA: Springer-Verlag New York, Inc., 1995, ISBN: 0-387-94559-8.

[8] A. J. Smola and B. Schölkopf, "A tutorial on support vector regression," *Statistics and Computing*, vol. 14, no. 3, pp. 199–222, 2004, ISSN: 1573-1375. DOI: `10.1023/B:STCO.0000035301.49549.88`. [Online]. Available: `http://dx.doi.org/10.1023/B:STCO.0000035301.49549.88`.

---

[4]`https://spark.apache.org/`

[9] J. Dean and S. Ghemawat, "MapReduce: Simplified data processing on large clusters," in *6th Symposium on Operating Systems Design and Implementation*, 2004, pp. 137–149.

[10] A. Verma, L. Cherkasova, and R. H. Campbell, "Aria: Automatic resource inference and allocation for mapreduce environments," in *Proceedings of the 8th ACM International Conference on Autonomic Computing*, ser. ICAC '11, Karlsruhe, Germany: ACM, 2011, pp. 235–244, ISBN: 978-1-4503-0607-2. DOI: 10.1145/1998582.1998637. [Online]. Available: http://doi.acm.org/10.1145/1998582.1998637.

[11] M. Malekimajd, D. Ardagna, M. Ciavotta, A. M. Rizzi, and M. Passacantando, "Optimal map reduce job capacity allocation in cloud systems," *SIGMETRICS Perform. Eval. Rev.*, vol. 42, no. 4, pp. 51–61, Jun. 2015, ISSN: 0163-5999. DOI: 10.1145/2788402.2788410. [Online]. Available: http://doi.acm.org/10.1145/2788402.2788410.

[12] E. Ipek, B. de Supinski, M. Schulz, and S. McKee, "An approach to performance prediction for parallel applications," English, in *Euro-Par 2005 Parallel Processing*, ser. Lecture Notes in Computer Science, J. Cunha and P. Medeiros, Eds., vol. 3648, Springer Berlin Heidelberg, 2005, pp. 196–205, ISBN: 978-3-540-28700-1. DOI: 10.1007/11549468_24. [Online]. Available: http://dx.doi.org/10.1007/11549468_24.

[13] S. Kundu, R. Rangaswami, K. Dutta, and M. Zhao, "Application performance modeling in a virtualized environment," in *High Performance Computer Architecture (HPCA), 2010 IEEE 16th International Symposium on*, Jan. 2010, pp. 1–10. DOI: 10.1109/HPCA.2010.5463058.

[14] J. Wildstrom, P. Stone, and E. Witchel, "Carve: A cognitive agent for resource value estimation," in *Autonomic Computing, 2008. ICAC '08. International Conference on*, Jun. 2008, pp. 182–191. DOI: 10.1109/ICAC.2008.27.

[15] S.-w. Liao, T.-H. Hung, D. Nguyen, C. Chou, C. Tu, and H. Zhou, "Machine learning-based prefetch optimization for data center applications," in *High Performance Computing Networking, Storage and Analysis, Proceedings of the Conference on*, Nov. 2009, pp. 1–10. DOI: 10.1145/1654059.1654116.

[16] E. İpek, S. A. McKee, R. Caruana, B. R. de Supinski, and M. Schulz, "Efficiently exploring architectural design spaces via predictive modeling," in *Proceedings of the 12th International Conference on Architectural Support for Programming Languages and Operating Systems*, ser. ASPLOS XII, San Jose, California, USA: ACM, 2006, pp. 195–206, ISBN: 1-59593-451-0. DOI: 10.1145/1168857.1168882. [Online]. Available: http://doi.acm.org/10.1145/1168857.1168882.

[17] B. C. Lee and D. M. Brooks, "Accurate and efficient regression modeling for microarchitectural performance and power prediction," in *Proceedings of the 12th International Conference on Architectural Support for Programming Languages and Operating Systems*, ser. ASPLOS XII, San Jose, California, USA: ACM, 2006, pp. 185–194, ISBN: 1-59593-451-0. DOI: `10.1145/1168857.1168881`. [Online]. Available: `http://doi.acm.org/10.1145/1168857.1168881`.

[18] J. L. Berral, Í. Goiri, R. Nou, F. Julià, J. Guitart, R. Gavaldà, and J. Torres, "Towards energy-aware scheduling in data centers using machine learning," in *Proceedings of the 1st International Conference on Energy-Efficient Computing and Networking*, ser. e-Energy '10, Passau, Germany: ACM, 2010, pp. 215–224, ISBN: 978-1-4503-0042-1. DOI: `10.1145/1791314.1791349`. [Online]. Available: `http://doi.acm.org/10.1145/1791314.1791349`.

[19] X. Yang and J. Sun, "An analytical performance model of MapReduce," in *IEEE International Conference on Cloud Computing and Intelligence Systems*, (Beijing), 2011, pp. 306–310, ISBN: 978-1-61284-203-5. DOI: `10.1109/CCIS.2011.6045080`.

[20] X. Cui, X. Lin, C. Hu, R. Zhang, and C. Wang, "Modeling the performance of MapReduce under resource contentions and task failures," in *IEEE Fifth International Conference on Cloud Computing Technology and Science*, (Bristol), 2013, pp. 158–163. DOI: `10.1109/CloudCom.2013.28`.

[21] Z. Zhang, L. Cherkasova, and B. T. Loo, "Performance modeling of MapReduce Jobs in heterogeneous Cloud environments," in *IEEE Sixth International Conference on Cloud Computing*, (Santa Clara, CA), 2013, pp. 839–846, ISBN: 978-0-7695-5028-2. DOI: `10.1109/CLOUD.2013.107`.

[22] N. Yigitbasi, T. L. Willke, G. Liao, and D. Epema, "Towards machine learning-based auto-tuning of MapReduce," in *IEEE 21st International Symposium on Modeling, Analysis & Simulation of Computer and Telecommunication Systems*, (San Francisco, CA), 2013, pp. 11–20. DOI: `10.1109/MASCOTS.2013.9`.

[23] P. Lama and X. Zhou, "Aroma: Automated resource allocation and configuration of MapReduce environment in the Cloud," in *Proceedings of the 9th International Conference on Autonomic Computing*, (San Jose, California, USA), ACM, 2012, pp. 63–72, ISBN: 978-1-4503-1520-3. DOI: `10.1145/2371536.2371547`.

[24] A. D. Popescu, A. Balmin, V. Ercegovac, and A. Ailamaki, "Predict: Towards predicting the runtime of large scale iterative analytics," *PVLDB*, vol. 6, no. 14, pp. 1678–1689, 2013. [Online]. Available: `http://www.vldb.org/pvldb/vol6/p1678-popescu.pdf`.

[25] D. Didona and P. Romano, "On bootstrapping machine learning performance predictors via analytical models," *CoRR*, vol. abs/1410.5102, 2014. [Online]. Available: `http://arxiv.org/abs/1410.5102`.